\documentclass[journal=acac3,manuscript=article]{achemso}
\usepackage{amssymb}

\author{T.V. Teperik$^{1}$, P. Nordlander$^{2}$,
J. Aizpurua$^{3}$ and A. G. Borisov$^{4}$}
\affiliation{$^1$ Institut d'Electronique Fondamentale,
UMR 8622 CNRS-Universit\'{e} Paris-Sud,
        \\ B\^atiment 220, 91405 Orsay Cedex, France. \\
$^2$ Department of Physics and Astronomy, \\
M.S. 61 Laboratory for Nanophotonics, Rice University,
\\ Houston, Texas 77251-1892, USA. \\
$^3$ Material Physics Center CSIC-UPV/EHU and \\
Donostia International Physics Center DIPC, \\
Paseo Manuel de Lardizabal 5 20018, Donostia-San Sebast\'{\i}an, Spain\\
$^4$ Institut des Sciences Mol\'{e}culaires d'Orsay,
UMR 8214 CNRS-Universit\'{e} Paris-Sud,
        \\ B\^atiment 351, 91405 Orsay Cedex, France.
}
\email{andrei.borissov@u-psud.fr}

\title
{Quantum Plasmonics: Nonlocal effects in coupled nanowire dimer.}

\begin{document}

\begin{abstract}
The optical response of a coupled nanowire dimer is studied
using a fully quantum mechanical approach. The translational
invariance of the system allows to apply the time--dependent
density functional theory for the plasmonic dimer with the largest
size considered so far in quantum calculations. Detailed comparisons
with results from classical electromagnetic calculations
based on local and non local hydrodynamic response, as well as
with results of the recently developed quantum corrected model is
performed. We show that electron tunneling and dynamical
screening are the major nonlocal quantum effects determining the
plasmonic modes and field enhancement in the system.
Account for the electron tunneling at small junction sizes allows
semi-quantitative description of quantum results within classical
framework. We also discuss the shortcomings of classical treatments
using non-local dielectric permittivities based on hydrodynamic
models. Finally, the implications of the actual
position of the screening charge density for the plasmon ruler
applications are demonstrated.
\\
\end{abstract}

In metal nanoparticles, collective excitation of the valence electrons
induced by an incident electromagnetic field, the localized plasmon,
leads to plethora of optical phenomena of significant current interest.
For instance, strong plasmonic enhancement of the local fields
\cite{Kelly,Javier,EnhRev,reviewNord,PasqualeetAl11ACSN}
opens a route to numerous practical applications, such as surface
enhanced Raman scattering (SERS) \cite{Raman,Raman1,Raman2,FazioetAl11ANSN},
optical nano-antennas \cite{Antenna0,Antenna} allowing, for example,
the control of radiation from single quantum
emitters \cite{Antenna1}, or generation of extreme ultraviolet
pulses by non-linear high harmonic generation \cite{NL2_Kim}.
In nanoparticle assemblies, the hybridization of plasmonic modes
can serve for guiding of the propagating fields
\cite{Quinten1998,Maier2003}, as well as it offers a way for rational
engineering of desired optical response and local field
profile.\cite{reviewNord} Not only light harvesting and
sensing properties can be thus greatly improved, but also the specific
geometry dependence of the optical response can be used as
plasmon ruler to determine the arrangement and nanoscale
distances within chemical or biological species.
\cite{reviewNord,Gunnarsson2005,Jain2007,
Hill2012,Ben2012,Liu2011,Acimovic2009}

Significant advances in fabrication and manipulation techniques
allow nowadays for precise control of the geometry of the
structure. \cite{Acimovic2009,JulurietAl11ACSN,Arielly2011,
Kern2012,Duan2012,Taylor2011,NL1_Novotny,Nature,Scholl2013}
In particular, for nanoparticle assemblies the size of the gap
between the adjacent nanoparticles can be brought below one
nanometer so that the electron tunneling across the junction
becomes possible. Thus, the plasmonic devices enter the quantum
regime which represents a significant challenge for available
theoretical approaches. Indeed,
most of the descriptions of the optical response are based on
the solution of the classical Maxwell equations where the
nanoparticles are modeled with sharp surfaces and the quantum
nature of electrons forming the screening charge is neglected.
Such effects as experimentally observed appearance of the charge
transfer plasmon for conductively coupled particles prior
to direct contact \cite{Nature,Scholl2013}, or optical
rectification \cite{Arielly2011,Ward2010} can not be addressed
within a classical framework. As shown with recent
quantum calculations
\cite{Nature,Zuolaga09,Marinica2012,Esteban2012,ZuloagaetAl10ACSN}
the main ingredients missing in the classical model are
the spill out of conduction electrons outside the nanoparticle
surfaces, and the finite spatial profile of the plasmon-induced
screening charge.  The use of a nonlocal dielectric functions
\cite{Javier,JGA_NL,David2011,Schatz_NL,Schatz_NL1,
Ciraci2012,Dominguez2012,Fernandez2012,Toscano2012,Toscano2012_1}
can account for the latter effect, where the variation of
electron density is smooth rather than infinitely sharp, as
assumed in the classical local approach. However, the spill out
of electrons outside the nanoparticle surfaces and associated
tunneling across the narrow interparticle junctions requires
special treatment.

In this respect, the data available on the coupled nanoparticle
dimer is quite revealing. This is a prototypical
system for plasmonic nanoparticle coupling, which has been
extensively studied. \cite{Javier,EnhRev,reviewNord}
Theoretical studies based on the solution of the classical Maxwell
equations predict a discontinuous transition from the capacitively
to conductively coupled particles. For vanishing junction
size, the fields in the junction diverge and the plasmon modes
experience diverging red shift as a result of the interaction
between high charge densities induced at the opposite sides of
the junction. \cite{EnhRev,reviewNord,PasqualeetAl11ACSN,Raman,
Raman1,Raman2,Hao2004,Romero06,Jain10}
The charge transfer plasmon appears abruptly after the
conductive contact. \cite{Romero06,Kottmann2001}
The nonlocal calculations based on the hydrodynamic model
\cite{Javier,JGA_NL,David2011,Dominguez2012,Fernandez2012,
Toscano2012} have shown that because of the finite spatial
profile of the plasmon-induced screening charge, the fields
in the junction stay finite albeit large. The number
of plasmon resonances and their frequency shift is much reduced
compared to classical predictions. On the other hand, it
follows from the quantum treatments
\cite{Nature,Zuolaga09,Marinica2012,Esteban2012}
that for narrow junctions, electron tunneling can
short circuit the junction and quench the plasmon-induced
field enhancement. The nanoparticles appear conductively
connected prior to direct contact, and the transition between
the non-touching and conductive contact regimes is
continuous. In particular, the charge transfer plasmon associated
with interparticle charge transfer \cite{Contact1_Disc,Contact2,
Schnell2009,Hentschel2011,Banik2012} progressively emerges in
the optical response of the system, as has been fully confirmed
in recent experiments.\cite{Nature,Scholl2013} These quantum
effects could be reproduced with Quantum Corrected Model
(QCM) \cite{Esteban2012} that treats the junction between
the nanoparticles as an effective medium allowing to account
for the quantum effects within the classical local Maxwell
theory.\cite{Nature,Scholl2013}

Present situation is thus characterised by the experimental
ability to create the structures with sub-nanometer characteristic
scales and several sometimes controversial theoretical approaches
to address the optical properties of such a nanostructures.
While quantum calculations provide an \textit{a priori} exact
answer to the problem, to date the application of rigorous
quantum mechanical approaches to coupled plasmonic systems have
been limited to rather small systems and are extremely heavy
numerically. It is thus of considerable interest to have a bench
mark quantum results on possibly large realistic system of coupled
plasmonic particles allowing detailed comparison
between full quantum and macroscopic Maxwell theories.
This would allow to assess of the role of quantum mechanical
effects and of the possibility to account for these effects
in classical theory. In this article which concerns a strongly
coupled nanowire dimer, we offer such a comparison. We show
that the quantum mechanical results for the optical response
of the system can be quantitatively reproduced with simple
models. \textit{Local classical} description of
the system allows to account for the interparticle tunneling
through the narrow junction and for the actual location of
the dynamically induced screening charges.

\section{Model and computational aspects}
%
%
\begin{figure}[tbp]
\resizebox{10cm}{!}{
\includegraphics*[angle=0] {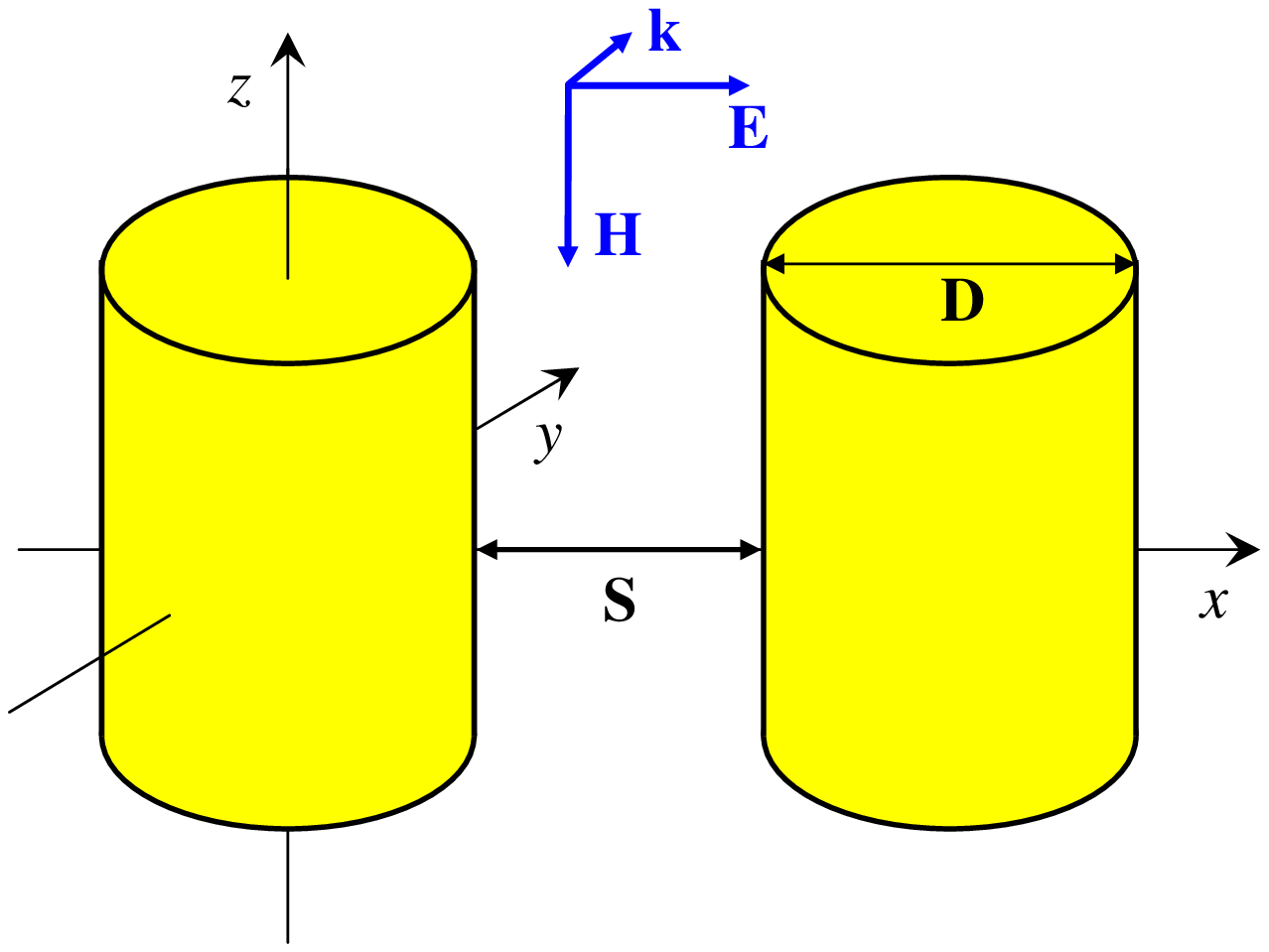} }
\caption{Sketch of geometry of the nanowire dimer.
Two identical cylindrical nanowires are infinite along $z$-axis
and have diameter $D$ of the circular cross-section in the
$(x,y)$- plane. The nanowires are separated by the junction of
width $S$. The incident radiation is linearly polarised with
electric field along $x$-axis.
}
\label{Fig:Sketch}
\end{figure}

The system considered in this work consists of two identical
nanowires in vacuum. The nanowires are infinitely extended
along $z$-axis as sketched in Fig.~\ref{Fig:Sketch}. Each
nanowire has a circular cross-section of diameter $D$. The
nanowires are separated by a variable distance $S$ and the
incident light is polarized along the inter-particle
axis $x$. Coupled nanowires and nanorods are often used in
SERS, nanoantenna, and plasmon guiding applications.
\cite{EnhRev,reviewNord,Antenna0,Antenna,
Liu2011,Kern2012,Berthelot2012,Manjavacas2009,Banholzer2008,
Alexander2010,Osberg2012} Thus, this system is well
characterized both from experimental and theoretical point of
view. Full numerical studies based on the solution of the
classical Maxwell equation have been performed \cite{Kern2012,
Kottmann2001,Kottmann2001_1,Halterman2005,Jain2006,
Funston2009,Tabor2009},
as well as semi-analytical studies based on transformation
optics.\cite{Dominguez2012,Fernandez2012,Lei2011}
The system has also been investigated using non-local
hydrodynamic description \cite{Dominguez2012,Fernandez2012,
Toscano2012,Toscano2012_1}. The high symmetry of the system
with its translational invariance along
the $z$-axis allows us to address on the full quantum level the
case of the cylinders with $D=9.8$ nm. To our knowledge this is
the largest size of the plasmonic dimer described so far within
the time dependent density functional theory (TDDFT) framework.

For the nanowires we adopt the cylindrical jellium model (JM).
Despite its simplicity, this model captures the collective
plasmonic modes of the conduction electrons and has demonstrated
its predictive power for quantum effects in nanoparticle
dimers \cite{Nature,Scholl2013}. While obviously not providing
chemical accuracy, the JM is well suited for the description of
the nonlocal effects due to conduction electrons such as
dynamical screening of external field and tunneling as we
discuss below. Along with the possibility to treat
relatively large system on a full quantum level, the JM model
allows for direct comparison between quantum and classical
electromagnetic theory results. Indeed, the physics underlying
the Drude model for the metal response as well as the
refinement of the Drude model introduced by the nonlocal
hydrodynamic model corresponds best to free electron metals.
For noble metals, such as silver and gold, the contribution
of the localised d-electrons to the screening would obscure
the comparison.

Within the JM, the ionic cores of the nanowire atoms are
represented with uniform background charge density
$n_{\rm{0}}=\left(\frac{4\pi}{3}~r_{\rm{s}}^3 \right)^{-1}$.
The screening radius $r_{\rm{s}}$ is set equal to $4~$a$_{\rm 0}$
(Bohr radius a$_{\rm 0}$=0.053~nm) corresponding to Na metal.
Sodium is a prototype system for which the JM performs particularly
well in description of the finite size non-local effects on optical
properties.\cite{Yannouleas1993} It should be emphasized that the
qualitative conclusions drawn in this work are robust and
independent of the particular choice of density parameter.
Each of nanowires is characterised by the number of electrons
$N_{\rm{e}}$ per unit length so that from the charge neutrality
the nanowire diameter is $D=2\sqrt{4 N_{\rm{e}} r_{\rm{s}}^3 /3}$.
The circle of diameter $D$ provides the position of the
jellium edge separating uniform positive background from
the vacuum. The jellium edge is located
at half a lattice constant $a$ ($a=4.23$~{\AA} for Na) in front
of the last atomic plane at the surface. We have performed
calculations for $N_{\rm{e}}=40$ and $N_{\rm{e}}=100$ with
$D_{40}=6.2$ nm and $D_{100}=9.8$ nm respectively.
The Fermi energy in both cases is at 2.9 eV below the vacuum
level.


The quantum calculations of the absorption cross-section are
based on the Kohn-Sham (KS) scheme of the time-dependent density
functional theory (TDDFT) \cite{TDDFT}. We use the adiabatic
local density approximation (ALDA) with the exchange-correlation
functional of Gunnarson and Lundqvist \cite{gunnarson76}.
Retardation effects can be neglected due to the small size
of the system. Extended description of present numerical
implementation can be found in Ref.~\cite{Marinica2012}.
Therefore we only discuss here specific aspects linked with
present work on interacting nanowires. In brief, the Kohn-Sham
orbitals of the ground state of interacting dimer
$\psi_j(x,y)$ are obtained from these of non-interacting
distant cylinders by adiabatically reducing the separation.
The $\psi_j(x,y)$ are discretized on the equidistant mesh in
cartesian coordinates, and the time dependent Kohn-Sham equations
\begin{equation}
i\frac{\partial \Psi_{j}(x,y,t)}{\partial t} ~=~\left(-\frac{\Delta}{2m}+V_{\rm{eff}}(x,y,t;[n])\right)
\Psi_{j}(x,y,t)
\label{Eq:TDSE}
\end{equation}
are then solved with initial conditions
$\Psi_{j}(x,y,t=0)=\psi_j(x,y)$.
In Eq.~(\ref{Eq:TDSE}) $m$ is the free electron mass, and
$\Delta=\partial^2/\partial x^2 + \partial^2/\partial y^2$.
The effective potential of the system $V_{\rm{eff}}$ depends
on the electron density $n(x,y,t)$ given by:
\begin{equation}
n(x,y,t)~=~ 2 \sum_{j \in occ.}
~\chi_j~\left| \Psi_{j}(x,y,t)\right|^{2}.
\label{Eq:Density}
\end{equation}
The sum runs over occupied states, the factor $2$ results
from the spin degeneracy, and $\chi_j$ is the number of
electronic states associated with $z$-motion along the nanowire.
\begin{equation}
\chi_j~=~ \frac{1}{\pi} \sqrt{2(E_F-E_j)}~,
\label{Eq:Degeneracy}
\end{equation}
where $E_F$ is the Fermi energy, and $E_j$ is the energy of the
$\psi_j(x,y)$ orbital.

The dipole absorption cross-section per unit length is 
calculated from the density
dynamics induced by impulsive perturbation as:
$\sigma_{abs}(\omega)=\frac{4\pi\omega}{c}
Im \left\{\alpha(\omega) \right\}$, with $\alpha(\omega)$ 
being the dipolar polarizability (per unit length) of the 
system, and $c$ - speed of light in vacuum.
In order to analyse the temporal evolution of the system
corresponding to the excitation of the given
plasmonic resonance we perform the TDDFT calculations
for the nanowire dimer subjected to an incident
$x$-polarized laser pulse. Electric field of the pulse is given
by $E_{L}(t)=E_{0}
\exp \left[ -\left(\frac{t-T}{0.2 T}\right)^{2}\right]
\cos \Omega t$. The frequency of the pulse $\Omega$ is
set resonant with studied plasmonic mode. The duration of the
pulse $T$ (typically 50 fs) is large enough for the narrow
spectral width. The amplitude $E_{0}$ is any small number
to guarantee the linear response regime \cite{Marinica2012}.
The snapshots of the induced densities, currents
and fields presented lower in this paper are extracted
at $t \sim T/2$.

The classical electromagnetic calculations of the absorption
cross-section have been performed with Comsol Multiphysics package,
(version 4.2a, www.comsol.com). Local, nonlocal, and quantum
corrected model description of the dielectric properties of the
system were used. It is noteworthy that because of the small
size of the plasmonic dimer the absorption cross-section closely
correspond to extinction cross-section in this case.

Within the local classical approach the dielectric constant
of the nanowires is described with Drude model, which is a
good approximation for the free electron metal as sodium
considered in this work.
\begin{equation}
 \varepsilon(\omega)=1-\frac{\omega_p^2}{\omega (\omega+i\gamma)},
 \label{Eq:Drude}
\end{equation}
where $\omega_p$ is the bulk plasma frequency given by
$\omega_p=\sqrt{4 \pi n_{\rm{0}}/m}=5.89$~eV, and $\gamma$
accounts for the damping.

The quantum corrected model (QCM) allows to introduce the tunneling
effect into the classical Maxwell equations. We use the local approach
where the  nanowires are described with Drude model Eq.~\ref{Eq:Drude},
and the tunneling across the vacuum gap between the nanowires is
accounted for by filling the junction with an effective dielectric
medium \cite{Esteban2012}. It is described with the Drude model,
similar to Eq.~\ref{Eq:Drude}:
\begin{equation}
 \varepsilon_{\rm{eff}}(y,S,\omega)=
 1-\frac{\omega_p^2}{\omega (\omega+i\gamma_{\rm{eff}}(y,S))}.
 \label{Eq:DrudeQCM}
\end{equation}
The effective damping $\gamma_{\rm{eff}}$ models transition
from resistive (large $S$) to conductive (small $S$) character
of the junction. Thus, it acquires a dependence on the separation
distance $S$ and $y$-coordinate. For very large $S$ the QCM is
exactly equivalent to local classical approach. No tunneling is
possible and the vacuum gap limit is retrieved
with  $\varepsilon_{\rm{eff}}(y,S,\omega)=1$. In this work we use
exactly the same parametrization of $\varepsilon_{\rm{eff}}$ as in
the earlier publication, where the QCM is introduced and described
in great detail \cite{Esteban2012}.

Within the non-local hydrodynamic description the transverse
component of the permittivity tensor is given by Eq.~(\ref{Eq:Drude}),
and the longitudinal component acquires the wave vector
$k$ dependence:
\begin{equation}
\varepsilon_L(\omega)=
1-\frac{\omega_p^2}{\omega (\omega+i\gamma)-\beta^2 k^2}.
\end{equation}
The $\beta$ parameter is given by $\beta=\sqrt{3/5} v_F$.
\cite{Javier,JGA_NL,David2011,Dominguez2012,Fernandez2012,
Toscano2012} In the present case, the Fermi velocity of
conduction electrons
$v_F=\sqrt[3]{3 \pi^2 n_{\rm{0}}}/m=1.05~10^6$~m/s, 
resulting in $\beta=0.81~10^6$ m/s.
Provided transverse and the longitudinal components of the
permittivity tensor the absorption cross-section is calculated
with numerical approach as recently implemented by G. Toscano
and coworkers within
the Comsol Multiphysics package.\cite{Toscano2012}

\section{Results and discussion}
\subsection{Individual Nanowire}
%
%
%
\begin{figure}[tbp]
\resizebox{16cm}{!}{
\includegraphics*[angle=270] {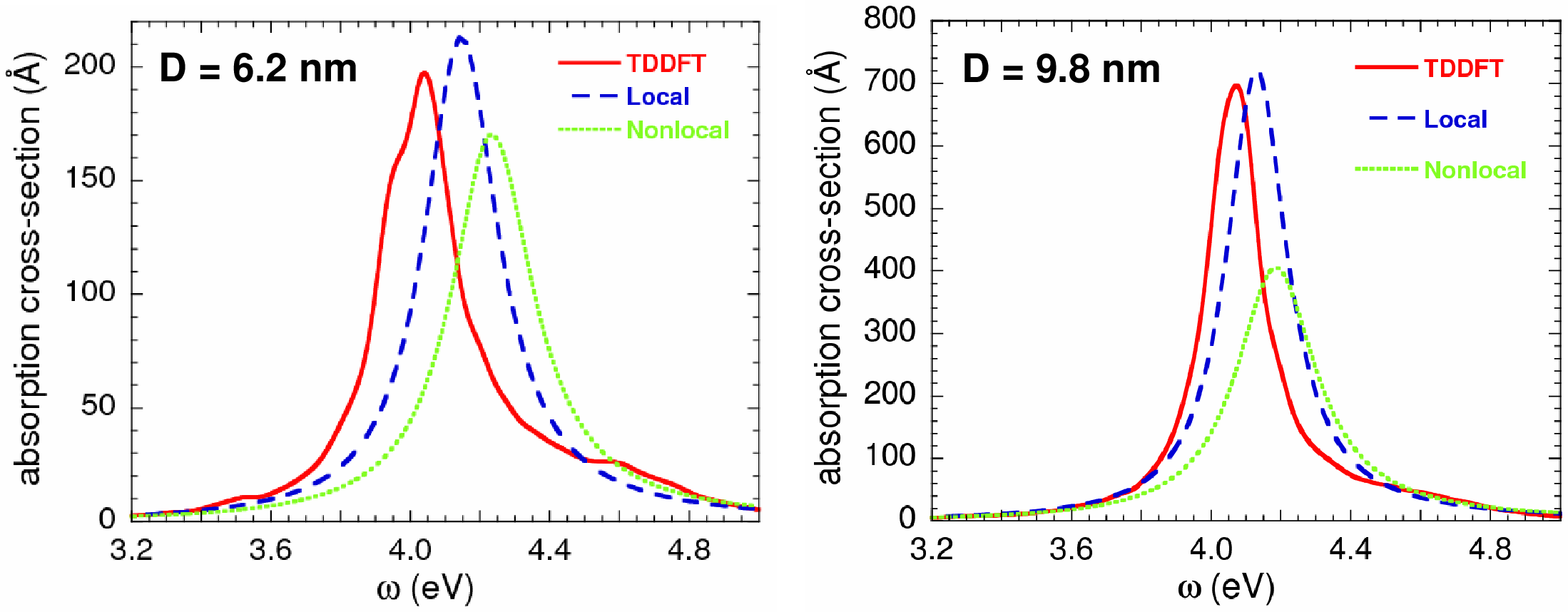} }
\caption{Absorption cross section $\sigma$ for the single jellium
nanowire of the diameter $D=6.2$ nm (left) and $D=9.8$ nm (right).
Results are shown as function of the frequency $\omega$ of the
incident radiation. The incoming field is the $x$-polarised plane
wave. The TDDFT calculations are compared with results of the
classical electromagnetic calculations using local (Drude) and
nonlocal hydrodynamic response. (See the legend for definition of
differnt symbols used in the figure).
}
\label{Fig:Single}
\end{figure}

We first characterize the linear optical response of
individual nanowire. In Fig.~\ref{Fig:Single} we show the dipole
absorption cross-section $\sigma$ calculated with full quantum
TDDFT method and with classical local Drude and non-local
hydrodynamic approaches. The classical local approach with the
Drude model given by Eq.~\ref{Eq:Drude} for the permittivity
of the nanowire leads to the dipole resonance located
at surface plasmon frequency
$\omega_{\rm{sp}}=\omega_p/\sqrt{2}=4.16$~eV irrespectively of
the nanowire diameter. We find that the absolute value of
$\sigma$ and the width of the resonance as calculated with TDDFT
is best reproduced with damping parameter $\gamma=0.275$~eV for
the $D=6.2$~nm nanowire, and $\gamma=0.247$~eV for the $D=9.8$~nm
nanowire. The smaller damping obtained for larger nanowire
indicates smaller decay due to surface scattering. Indeed,
the surface scattering contribution to the damping can be
expressed as $\gamma_S= A v_F/R_{\rm{cl}}$ \cite{Hovel1993},
where $R_{\rm{cl}}=D/2$ is the cylinder radius, and $A$ is the
system-dependent parameter. The calculated change of the damping
is consistent with $A=0.5$ in accord with available
experimental and theoretical results for alkali
clusters.\cite{AppelPenn1983,Klein1998} Obviously, through the
width of the single wire resonance the quantum size
effects would affect the optical response of the nanowire
dimer.\cite{Osa2012}

The dipole resonance calculated with TDDFT is
\textit{red shifted} from the classical $\omega_{\rm{sp}}$
value but approaches $\omega_{\rm{sp}}$
for increasing size of the system. For the $D=6.2$~nm
nanowire, the resonance is at $\omega=4.027$~eV, and it is
at $\omega=4.072$~eV for the $D=9.8$~nm nanowire. Thus,
present TDDFT results for cylindrical nanowire show the same
trends as studied in detail for spherical alkali metal
clusters.\cite{Yannouleas1993,Parks1989,Borggreen1993,Reiners1995}
The non-local effect at the origin of the red shift of the
plasmon resonance is linked with dynamical screening of the
fields by conduction electrons.
The dipole resonance frequency can be found
from \cite{Borggreen1993,Reiners1995,Apell1982,Liebsch1993}:
\begin{equation}
\omega/\omega_{\rm{sp}}=1-Re[d(\omega_{\rm{sp}})]/R_{\rm{cl}}+
O(R_{\rm{cl}}^{-2}).
\label{Eq:RedShift}
\end{equation}
The real part of Feibelman's parameter $Re[d(\omega)]$ gives
the position of the centroid of the induced charge density
\cite{Feibelman1982,Appel1984, Liebsch1987}.
When measured from the jellium edge, $Re[d(\omega_{\rm{sp}})]$
is positive for alkali metals,
i.e. the screening charge is shifted into the vacuum because
of the spill out of the conduction electrons outside the metal.
From the calculated resonance frequency we obtain
$Re[d(\omega_{\rm{sp}})]=1$~{\AA} in
agreement with earlier calculations and experiment
\cite{Reiners1995,Feibelman1982,Liebsch1987}.
Importantly, for the noble metals such as Au and Ag the
final size effects lead to the \textit{blue shift} of
the dipole plasmon resonance \cite{Tigges1993,Scholl2013}.
The difference between alkali and noble metals is the localised
d-electron contribution to the total screening in the latter
case.\cite{Liebsch1993,Appel1984,Tigges1993,Liebsch1995,
Serra1997} When the d-electron
contribution is accounted for, $Re[d(\omega)]$ turns negative
indicating that screening charge is predominantly induced
inside metal.\cite{Liebsch1993,Tigges1993,Liebsch1995}

For the small diameter nanowire $D=6.2$~nm, finite size
effects appear in the TDDFT calculations not only in the form
of red shifts and line width changes, but also as clearly
observable structures in the frequency dependence of the
absorption cross section. These structures arise from the
strong coupling of collective plasmon- and single electron-hole
excitations. For large diameter nanowires, finite size effects
become smaller and the plasmon resonance is much better defined.
The TDDFT results also show shallow resonance in the optical
absorption cross section at $\omega=4.6$~eV associated
with a multipole plasmon (MP)
\cite{Tsuei1990,Bennett1970,Pitarke2007}.
The resonant frequency obtained with TDDFT is in good
agreement with experimental data.\cite{Tsuei1990,Pitarke2007}
The density oscillations of the MP proceed within the layer
of the spilled out charge as obviously can not be captured
within the local classical theory.

Finally, the green dotted lines in Fig.~\ref{Fig:Single}
show the absorption cross-section calculated with the
classical non-local hydrodynamic model 
(NLHD) \cite{Toscano2012}. In contrast to the
TDDFT result and experimental data on alkali
clusters \cite{Borggreen1993,Reiners1995}, the NLHD model
predicts a blue shift of the dipole plasmon frequency from
$\omega_{\rm{sp}}$ value. The reason for this is that
independent of the metal, the plasmon-induced charges in
NLHD are localized within the layer of the thickness
$\beta/\omega_p$ below the metal
surface \cite{Fernandez2012}. Thus, the effective
$Re[d_{\rm{NLHD}}(\omega)]$ is always
negative leading to the blue shift of the localised
plasmon \cite{Appel1984} in contradiction with the spill
out effect known for alkali metals. However, negative
$Re[d_{\rm{NLHD}}(\omega)]$ inherent to NLHD description of
conduction electrons and so the blue shift of the localised
plasmon qualitatively coincides with full quantum result
determined by the $d$-electron screening for the noble
metals.\cite{Liebsch1993,Appel1984,
Tigges1993,Liebsch1995,Serra1997,Feibelman1982,
Liebsch1987,Tsuei1990} 

%
%
\begin{figure}[tbp]
\resizebox{16cm}{!}{
\includegraphics*[angle=270] {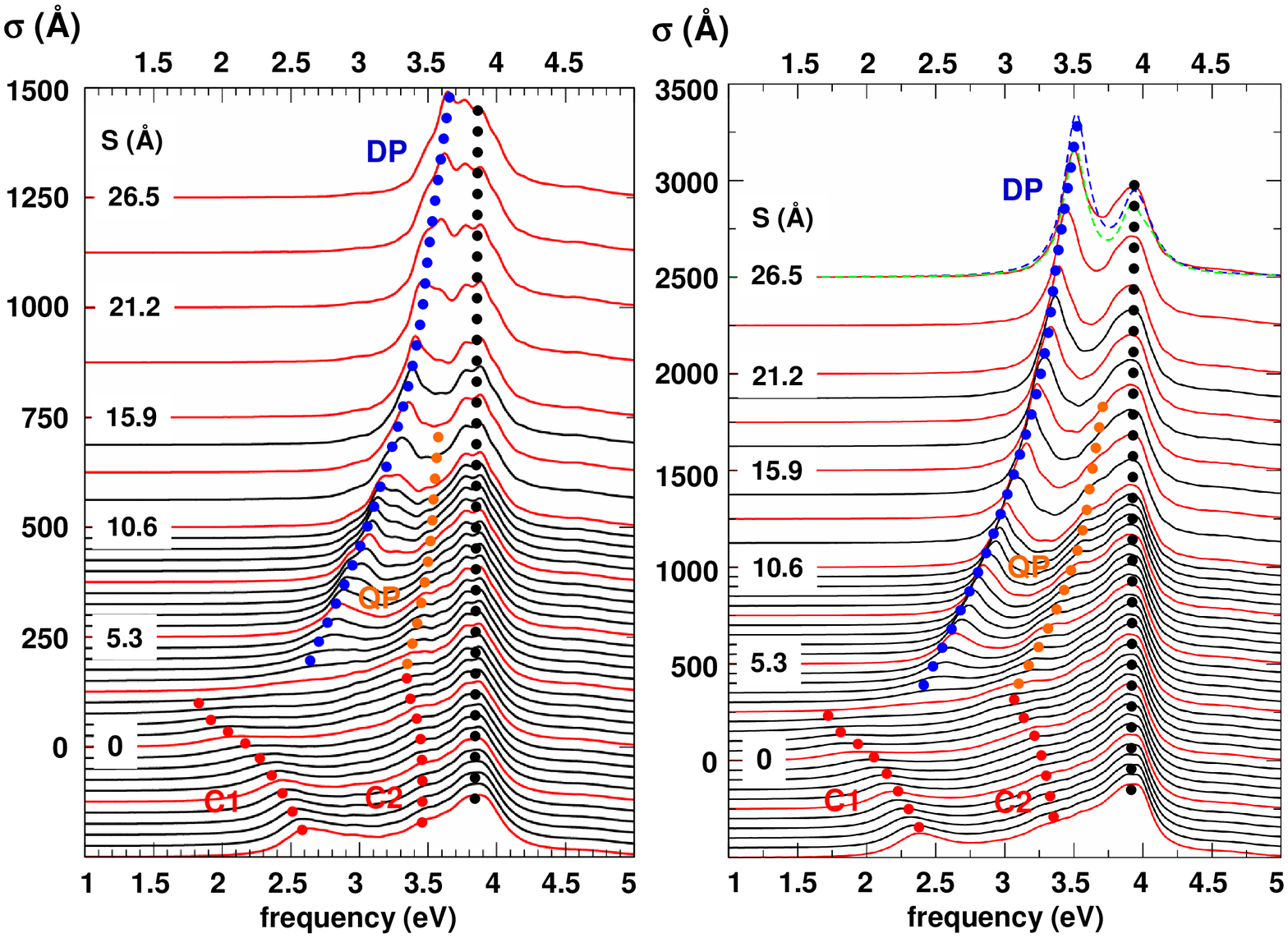} }
\caption{
Waterfall plot of the dipole absorption cross-section
for the nanowire dimer in vacuum. The dimer consists of two
Na nanowires of diameter $D=6.2$ nm (left) and $D=9.8$ nm
(right) separated by variable distance $S$. The incoming
field is the $x$-polarised plane wave. The centers
of the particles are at $x=\pm (D+S)/2)$, and $S$ is negative
for overlapping cylinders. $S=-D$ would correspond to the limit
of a single cylinder. TDDFT results are given as function of the
frequency $\omega$ of the incident radiation for different
separations $S$. For clarity a vertical shift proportional to
the separation distance is introduced for each absorption
spectrum. The red curves are used
each $5~$a$_{\rm 0}\approx 2.65$ {\AA} of $S$-change. These
are labeled with corresponding $S$-values
each $10~$a$_{\rm 0}\approx 5.3$ {\AA} of $S$-change.
The plasmonic modes responsible for the peaks in the absorption
cross-section are labelled. DP stands for the Bonding Dipole
Plasmon, QP for Bonding Quadrupole Plasmon, and C1 for the
lowest (dipole) Charge Transfer Plasmon, C2 for the higher
energy Charge Transfer Plasmon. Vertical black dotted line
shows the degenerate plasmon peak located close to
$\omega_{\rm{sp}}$. On the right panel the blue (green) dotted curve
at $S=26.5$ {\AA} shows results of the classical Drude
(NLHD) calculations with adjusted parameters.
Further details are give in the main text of the paper.
}
\label{Fig:LinearResponse}
\end{figure}

\subsection{Coupled Nanowires}

In Fig.~\ref{Fig:LinearResponse} we show waterfall plots
of the dipole absorption cross-section of the $D=6.2$~nm and
$D=9.8$~nm nanowire dimers calculated with TDDFT.
The results are presented as function of the frequency of the
incident radiation for different widths $S$ of the junction.
The calculations have been performed for both positive and 
negative $S$, where the latter means a geometrical overlap 
of the nanowires. The $S=0$ case corresponds to the kissing 
cylinders \cite{Dominguez2012,Fernandez2012} with
touching jellium edges where the distance between the topmost
atomic planes equals to the interlayer spacing, i.e.
a continuous solid is formed at the contact point.

\begin{figure}[tbp]
\resizebox{16cm}{!}{
\includegraphics*[angle=270] {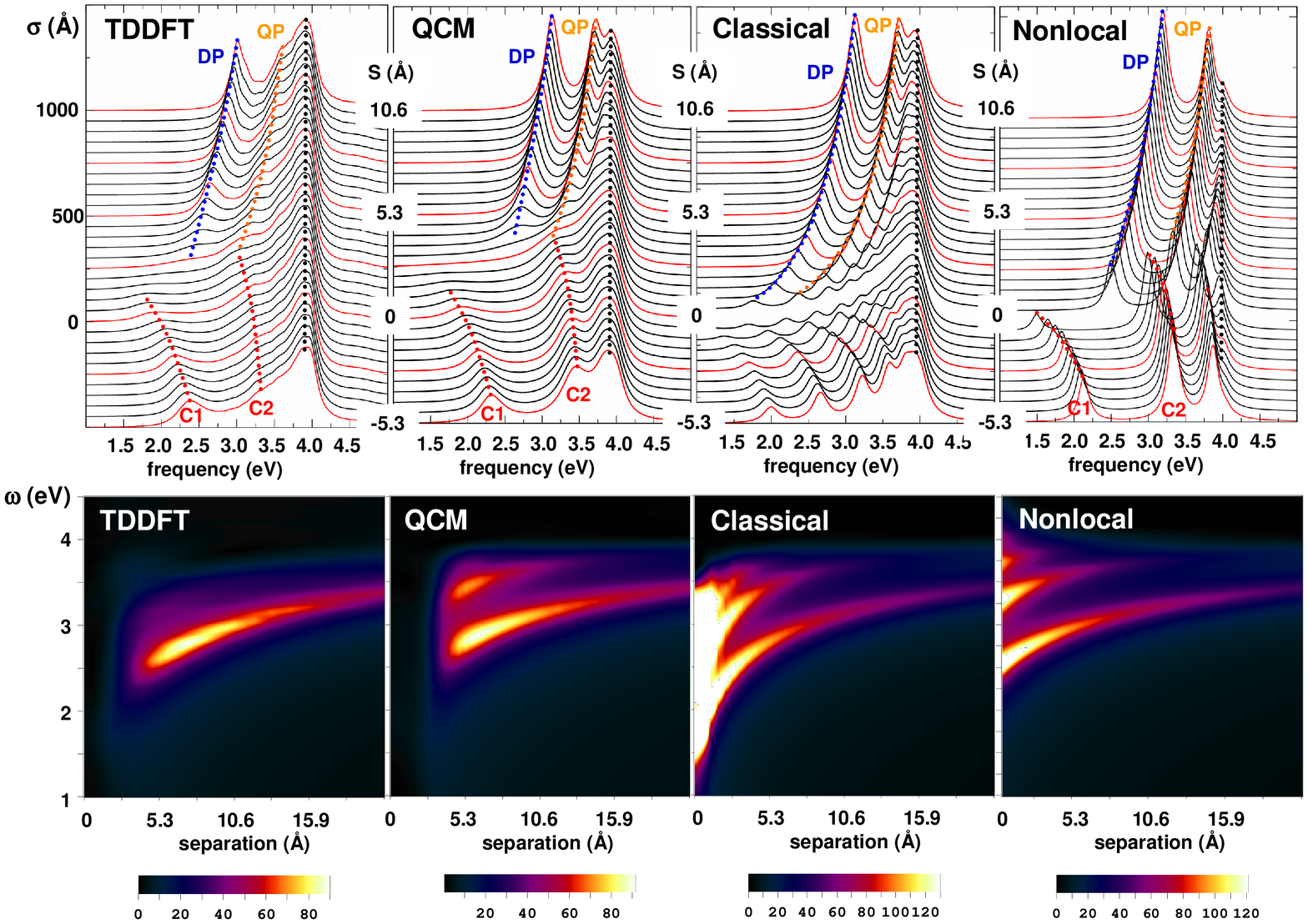} }
\caption{
Comparison of the optical properties of a nanowire dimer
obtained with the full TDDFT calculations, with the quantum
corrected model (QCM), with a classical Drude electromagnetic
calculations (Classical), and with calculations based on the
nonlocal hydrodynamic model (Nonlocal). The dimer consists of
two $D=9.8$ nm Na nanowires in vacuum. The incoming plane
wave is polarized along the dimer axis $x$.
Upper panels: Waterfall plots of the dipole absorption
cross-section as function of the width of the junction $S$.
Red curves correspond to separation distances of
$S=-5.3$ {\AA}, $-2.65$ {\AA}, $0$ {\AA}, $2.65$ {\AA},
$5.3$ {\AA}, $7.95$ {\AA}, and $10.6$ {\AA}. For further
details see caption of Fig.\ref{Fig:LinearResponse}
Lower panels: Color plots of the local field enhancement at
the center of the junction for positive separations. Results
are shown as function of the frequency $\omega$ of the incident
radiation and separation distance $S$. The color code is
explained at the bottom of the corresponding panels.
}
\label{Fig:Models}
\end{figure}

For large positive separations the non-local effects
including electron tunneling across the junction are small,
and the TDDFT results agree with earlier
classical calculations \cite{Fernandez2012,Kottmann2001,
Kottmann2001_1,Halterman2005,Fernandez2012}
(See also Fig.~\ref{Fig:Models}). At $S=26.5$~{\AA} the
absorption spectrum is dominated by two resonant structures.
First is the bonding dipole plasmon (DP)
indicated with blue dotted line in Figure. It is formed from
hybridization of the dipole plasmon modes of the individual
nanowires. The second resonance indicated with black dotted
line is formed by bonding quadrupole (QP) plasmon degenerated 
with higher order modes. It is slightly red shifted with
respect to the surface plasmon frequency $\omega_{\rm{sp}}$.
At $S \rightarrow \infty$, the DP merges into the degenerate
mode and the spectrum evolves into that of the individual
nanowire. As the junction width $S$ decreases, the
DP shifts to lower frequencies because of the attractive
interaction between the charges of opposite sign across the
junction. Along with red shift of the DP, the QP (orange
dotted line in Figure) splits from the degenerate mode, and
shows the red shift with decreasing $S$.

Despite the overall similarity of the results obtained with
smaller and larger diameter nanowires there are some
notable differences primarily caused by finite size effect.
The resonances are much better defined for the larger
$D=9.8$~nm nanowire. In this case the width of the
resonances is smaller and the structures due to the
interaction between the plasmon and electron-hole pair
excitations disappear. For $D=6.2$~nm nanowire, this
interaction can even split the single plasmon peak into
several lines. The red shift of the DP and QP modes with
decreasing $S$ is more pronounced for the $D=9.8$~nm
nanowire dimer. This is because of the scale ($S/D$)
invariance of the plasmonic properties of nanostructure
in the quasistatic limit.

For junction widths below $\sim 7$~{\AA}, electron
tunneling across the junction becomes important.
The results obtained here for the nanowire dimer have much
in common with those reported in quantum studies of
touching spherical nanospheres \cite{Zuolaga09,Marinica2012,
Esteban2012}, and in recent experiments \cite{Nature,Scholl2013}.
The DP resonance progressively disappears and the charge
transfer plasmon mode (C1) emerges prior to the direct contact
between the nanoparticles. C1 appears as a broad shallow
low-frequency peak at positive $S \simeq 1.5$~{\AA}, and
evolves into a well--defined resonance at $S<0$. Similarly,
because of electron tunneling, the QP mode continuously
evolves into a higher order charge transfer plasmon mode C2
before direct contact between the nanowires. Thus, already
at positive $S$ the nanowires appear
conductively connected showing characteristic charge
transfer plasmon modes \cite{Romero06,Contact1_Disc,
Contact2,Schnell2009,Hentschel2011}.
For a dimer with well established conductive contact, the
C1 and C2 modes experience a blue shift as also reported
in classical calculations \cite{Romero06,Lei2011}.

%
%
\begin{figure}[tbp]
\resizebox{16cm}{!}{
\includegraphics*[angle=270] {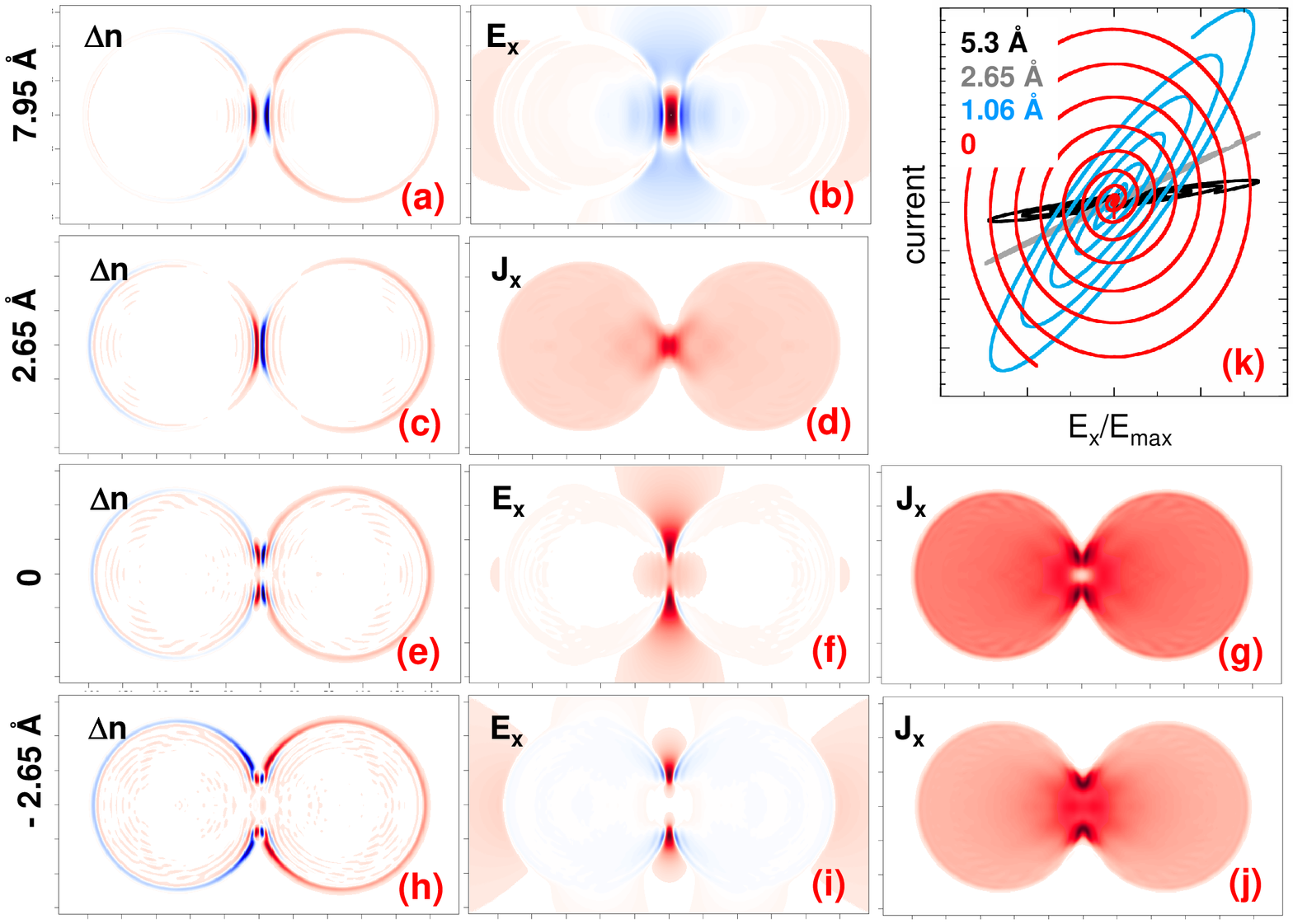} }
\caption{Panels (a)-(k) Detailed analysis of the plasmon dynamics
in the coupled Na nanowires of diameter $D=9.8$ nm. The incident
$x$-polarised laser pulse is at resonance with lowest (dipole
plasmon mode of the system).
Panels (a)-(j) present the snapshots
of the induced charge density $\Delta n$, current density $J_x$,
and field $E_x$ for different junction widths $S$ as
indicated to the left of each row. The induced currents and fields
are measured along interparticle $x$-axis. Positive (negative)
values correspond to red (blue) color code. The induced densities
are shown at the instant of time corresponding to the maximum
dipole moment of the dimer. The induced currents and fields are
shown at the instants of time when the induced fields in the
junction reach maximum.
Panel (k): Conductivity analysis. The current $J_x$ measured
on $x$-axis in the middle of the junction is plotted as a
function of the normalized electric field at the same position.
Different colors correspond to different separations $S$ as
explained in the insert.
}
\label{Fig:Snapshots}
\end{figure}

To obtain further insights into the effect of the tunneling
through the junction we have calculated the electron density
dynamics in the plasmonic dimer subjected to an incident
$x$-polarized laser pulse resonant with lower (DP or C1) plasmonic
modes.
The panels (a)--(j) of Fig.~\ref{Fig:Snapshots}
show snapshots of the induced charge density $\Delta n$, and
$x$-component of the current density $J_x$, and field
$E_x$ for different junction widths $S$. Results are given
for the case of $D=9.8$ nm nanowires. The induced
densities are shown at the instant of time corresponding
to the maximum dipole moment of the dimer. The currents and
fields are shown at the instant of time corresponding to the
maximum induced field in the junction.

For large separation $S=7.95$~{\AA} the maximum induced
dipole corresponds to the in--phase dipole polarisation
of each nanocylinder expected for the DP mode. High charge
densities are induced at the surfaces facing the junction
resulting in large electric field enhancement
$|E_0/E_{in}| \sim 80$.  Here, $E_{in}$ is the
amplitude of the incident field and $E_0$ is the amplitude
of the field measured at the $x$-axis in the middle of
the junction. The structure of the induced charges and fields
are similar to previous classical results for coupled
cylinders \cite{Fernandez2012,Kottmann2001,Kottmann2001_1,
Halterman2005} and also resembles much to the case of
metal sphere dimer \cite{Romero06}. The probability of
tunnelling between nanowires is negligible and no current
flows across the junction. Note that the maximum dipole
polarisation corresponds to the instant of time when the
maximum charge separation has occurred and the currents
inside nanoparticles are minimal.
For the reduced width $S=2.65$~{\AA}, very
similar profiles for induced densities and fields are obtained.
Therefore, we do not show $E_x$ but focus on the induced
current. The junction width is now sufficiently small to allow
weak electron tunnelling between the nanoparticles. Large
optical field in the junction acts as large BIAS in the 
scanning tunneling microscope causing tunneling current 
across the junction.\cite{Esteban2012} Thus, the junction
shows the \textit{resistive} character with maximum
current between nanowires reached at the maximum
field and consequently at maximum induced dipole.

Further reduction of the separation $S$ increases
the tunneling probability and short circuits the junction.
When conductive contact is formed, the DP mode disappears
and the C1 mode emerges in the absorption spectrum.
Panels (e) and (h) Fig.~\ref{Fig:Snapshots} show the
induced charges for $S=0$ and $S=-2.65$~{\AA}, respectively.
The results are very similar, i.e. already for the $S=0$ case
of kissing cylinders the charge transfer plasmon mode is well
developed (see also Fig.~\ref{Fig:LinearResponse}, and
Fig.~\ref{Fig:Models}). The maximum dipole moment of the
dimer corresponds to the oppositely charged nanowires. 
Maximum currents and fields [panels (f), (g), (i), (j)] are 
reached when the total dipole moment of the system is minimum
(compare with $S=2.65$~{\AA} case). Precisely, this is large
current flowing trough the entire system that builds the
dipole polarisation with opposite charges at left and right
nanowires consistent with resonant excitation of the C1
mode. The field enhancement is about 30 for both $S=0$ and
$S=-2.65$~{\AA} separations. However, the fields are 
screened at the center of the junction, and the maximum fields 
are located at its sides. This is similar to the classical 
result for the overlapping cylinders.\cite{Lei2011}

Panel (k) of Fig.~\ref{Fig:Snapshots} provides further analysis
for the evolution from resistive to conductive character of the
junction with decreasing $S$. The current $J_x(t)$ on the $x$-axis
at the middle of the junction between the nanowires is shown as a
function of the field $E_x(t)$ at the same location. At large
separations, the linear relation $J_x(t)= g E_x(t)$ between the
current and the local field shows that the junction is resistive.
The increase of the slope $g$ when $S$ is reduced from $S=5.3$~{\AA}
to $S=2.65$~{\AA} is because of the increase of the tunneling
probability. For $S=1.06$~{\AA} and $S=0$~{\AA} the $E_x(t)$ and
$J_x(t)$ acquire a relative phase. Since the field envelope of
the incident pulse grows in time for the time interval shown in
Fig.~\ref{Fig:Snapshots}(k), the clock-wise rotation of the
$J_x(t)[E_x(t)]$ curve implies that the current is retarded with
respect to the field. The junction becomes conductive which is
particularly apparent for $S=0$.

%
%

The TDDFT results show that for small junction widths
the optical response is determined by electron tunneling.
For large width of the junction the tunneling is
negligible. Each cylinder responds on the self
consistent field created by the incident radiation and the neighbour.
The effects of nonlocal screening discussed for the single cylinder
should \textit{a priori} influence the optical response of the dimer
as suggested with NLHD calculations
\cite{Dominguez2012,Fernandez2012, Toscano2012,Toscano2012_1} and
as inherently build in the TDDFT results. While the effect of
tunneling is clearly seen with TDDFT calculations, revealing the
role of the nonlocal screening requires comparison with classical
results, in particular for the energies of the plasmonic
modes. In this respect, present system offers a unique opportunity
to test the validity of different classical models against
full quantum results for relatively large system, with fully
developed plasmonic modes, and with dielectric properties well
described within free electron approximation. The rest
of the paper is devoted to quantum vs classical comparison with
particular focus on possibility to account for quantum effects
within classical approach.

\subsection{Quantum vs classcal approaches}

Fig.~\ref{Fig:Models} shows the present TDDFT results, the QCM
results, results from classical calculations based on the
Drude description of Na dielectric function, and results of the
NLHD calculations. Waterfall plots of the
absorption cross-section (upper panels) and contour plots of the
field enhancement on the $x$-axis at the middle of the junction
are presented as function of the frequency of the incoming plane
wave and of the junction width $S$.  Since for the individual 
nanowire TDDFT gives a red shift of the dipole plasmon
resonance from the classical $\omega_{\rm{sp}}$ position, while
NLHD leads to a blue shift (Fig.~\ref{Fig:Single}), we adjust
the parameters of different models. For the sake of comparison
of the junction width dependence we enforce an
agreement between the TDDFT, QCM, classical Drude, and
NLHD at the largest $S$ calculated here with TDDFT, $S=26.5$~{\AA}
(see Fig.~\ref{Fig:LinearResponse}). For the $D=9.8$ nm nanowire
dimer we have used plasma frequency
$\omega_{p}=5.8$~eV and damping $\gamma=0.247$~eV in the QCM
and classical Drude calculations. Thus, only small correction to
the bulk plasma frequency (nominal value $5.89$~eV) is required
in this case. The NLHD model needs for much larger correction to
compensate for the blue shift inherent to this treatment. We
used $\omega_{p}=5.5$~eV and $\gamma=0.16$~eV. Adjustment of
the nonlocality parameter $\beta$ is also possible, but then the
model looses its predictive power. Indeed, as far adjustment
is limited to $\omega_{p}$ and $\gamma$ parameters, the finite
size effects and so the need for correction will disappear with
increasing size of the nanoobjects.

\begin{figure}[tbp]
\resizebox{16cm}{!}{
\includegraphics*[angle=270] {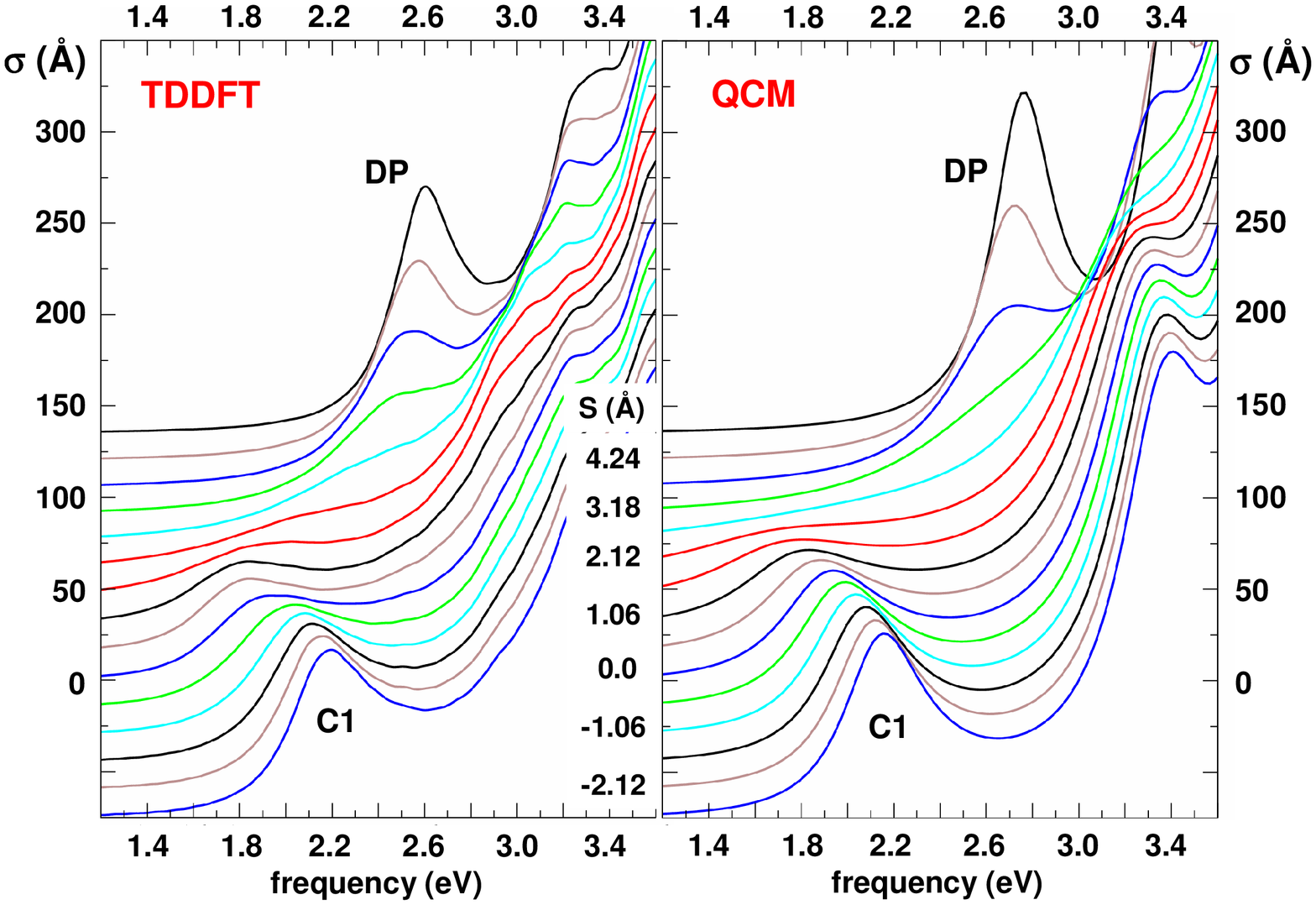} }
\caption{
Detailed comparison between TDDFT and QCM calculations.
The dipole absorption cross-section of the $D=9.8$ nm
Na nanowire dimer is shown for small separations $S$
corresponding to the strong tunneling regime and
transition from separated to conductively coupled
nanowires. The frequency range is zoomed at the transition
from the bonding dipole (DP) to the lowest charge
transfer (C1) plasmon.
Waterfall plots of the optical absorption cross-section are
shown for the separation distances changing from
$S=-2.65$ {\AA} (lowest blue line) to $S=4.77$ {\AA}
(upper black line) in steps of $1~a_0$ (0.53 {\AA}).
For further details see caption of
Fig.~\ref{Fig:LinearResponse}.
}
\label{Fig:ZoomCompare}
\end{figure}

The main features of the quantum results have been discussed
in connection with Fig.~\ref{Fig:LinearResponse} so here we
will focus on comparisons between different model approaches.
As follows from Fig.~\ref{Fig:Models}, the QCM does an
excellent job in describing the TDDFT results over the entire
range of separations $S$ addressed here.
The important features such as: the number of resonances;
their distance dependence; the transition from the separated
to conductively coupled regime are well reproduced.
In particular, in sheer contrast with classical theory
\cite{Romero06,Fernandez2012,Kottmann2001,Kottmann2001_1,Halterman2005}
the change of the spectrum at the moment of contact is
progressive and fields at the middle of the junction are quenched,
not diverging with decreasing junction size. Besides qualitative
aspects, we find that the TDDFT and QCM agree semi-quantitatively
as is further stressed in Fig.~\ref{Fig:ZoomCompare}. This figure
zooms into the most delicate interaction regime corresponding
to the transition from the separate
to overlapping nanowires.

By construction of QCM, it is equivalent to the classical 
local Drude description for large positive $S$ where 
tunneling is negligible. The good agreement with TDDFT data 
suggests that the pure local classical description is
reasonable for a large $S$. At the same time the classical 
description fails at small $S \sim 5$~{\AA}, i.e. typically 
at two lattice constants between surface atomic planes that 
define the junction \cite{Scholl2013}. The accumulation of 
charges on the opposite sides of the junction leads to 
exaggerated coupling between nanowires resulting in diverging 
fields and too large number of resonances. Similarly, for 
negative $S$, the sharp edges of the junction, which 
are otherwise smeared out by the electron tunneling also 
result in too many hybridized resonances \cite{Esteban2012}.

%
%

The NLHD description is free from the $S=0$ divergence
problem.\cite{Dominguez2012,Fernandez2012} The number of
resonances remains small and the fields in the middle of the
junction stay finite albeit too large. This can be easily
understood thanks to the elegant transformation optics
approach developed for interacting cylinders by
Fern\'{a}ndez-Dom\'{\i}nguez and coworkers \cite{Fernandez2012}.
At positive $S$ an analytic solution is found which depends
on the renormalised parameter
$\tilde{S} \simeq S+ 2 \delta$, where
$\delta \simeq \beta/\omega_p$ (1 {\AA} in the present case).
The physics behind this shift is the localisation of the
plasmon-induced charge below the surface as inherent for the
NLHD model. As we discussed for the single cylinder, in terms
of the dynamical screening theory, $\delta$ is equivalent to an
effective Feibelman parameter $\delta=-Re[d_{\rm{NLHD}}(\omega)]$.
Thus, even for $S=0$, the induced charges at the opposite sides
of the junction are actually separated by a finite
distance $2 \delta$.
However, since the tunneling is not accounted for, the NLHD
description fails to reproduce the quenching of the field
enhancement at the middle of the junction for small positive $S$.
The NLHD also fails to smoothen out the transition from
separated to overlapping regimes and gives an abrupt nonphysical
transition. The number of modes is smaller than in the classical
Drude description, but still larger than what is obtained in
TDDFT or QCM calculations.

Our final remark concerns the QP plasmon that evolves into
the C2 charge transfer mode for negative $S$. The associated
resonances are much less pronounced in the TDDFT cross-section
compared to model approaches. One possible reason is
finite size effect, where the the system size is not large
enough for the corresponding density oscillations to be
completely formed. However, the similarity between the TDDFT
results for $D=6.2$ nm and $D=9.8$ nm nanowire dimers as shown
in Fig.~\ref{Fig:LinearResponse} suggests that this quantum size
effect is small. We thus tentatively attribute the weaker
high order resonances as obtained in TDDFT to the effect of the
smearing of the induced surface charge densities. This
\textit{a priori} reduces the coupling between the dipole and
higher order modes and consequently the intensity of the
QP resonance in the dipole absorption cross-section.

\subsection{Dynamic screening}

%
%
\begin{figure}[tbp]
\resizebox{16cm}{!}{
\includegraphics*[angle=270] {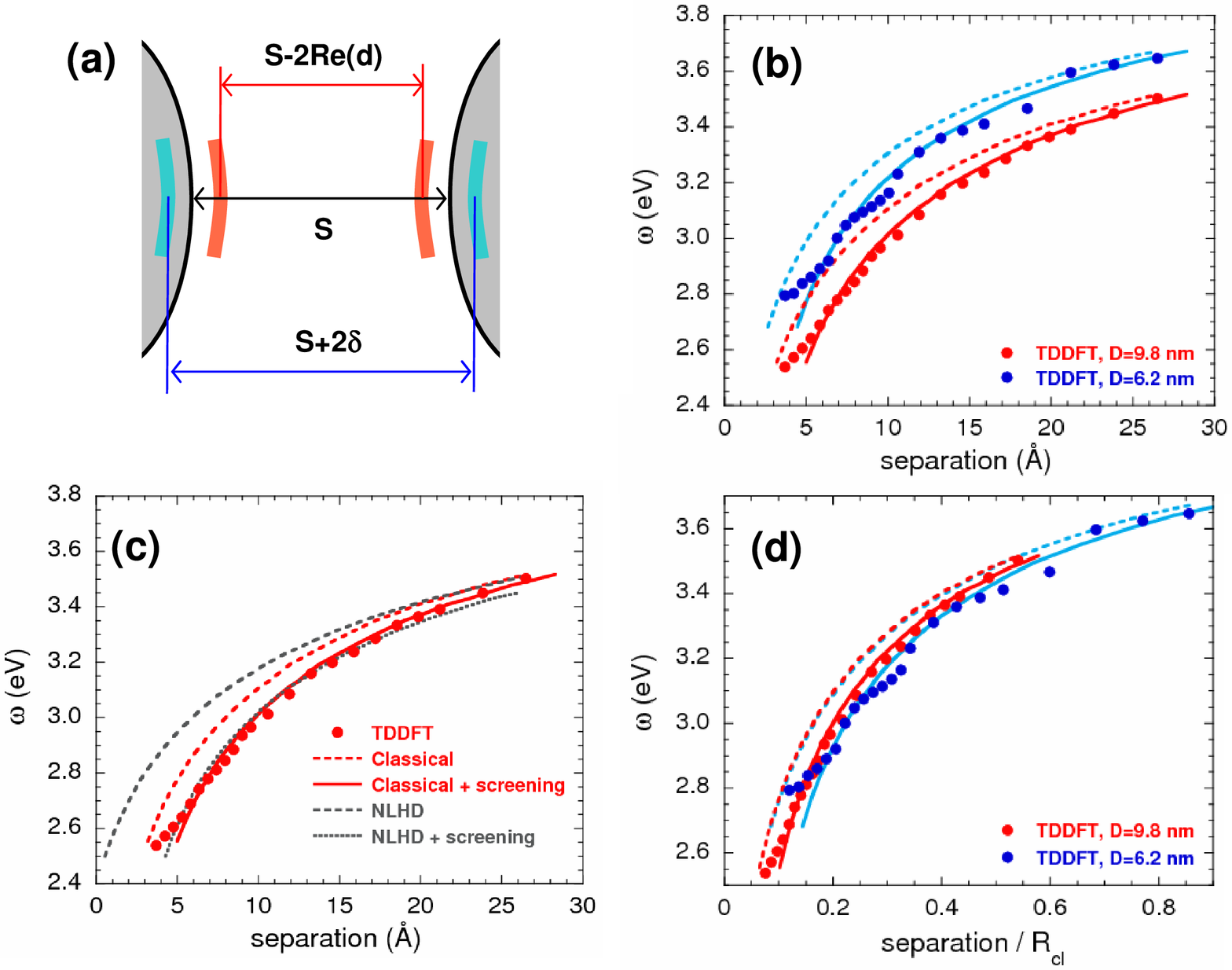} }
\caption{Dynamic screening. (a) Schematic representation of the
location of plasmon induced screening charges in the junction.
Within the local classical approach the screening charges are
at geometrical surfaces of the cylinders (here  equivalent to
the jellium edges) separated by the junction of width $S$. 
Within the TDDFT, the centroids of the screening charges 
(red areas) are located at $Re[d(\omega)]$ in front of the 
jellium edges and separated by $S-2 Re[d(\omega)]$.
In NLHD approach the centroids of the  screening charges
(blue areas) are located at distance
$\delta$ below the geometrical surface and
separated by $S+2 \delta$.
(b) Energy of the dipole plasmon resonance as function of the
junction width $S$. Dots: TDDFT results obtained for nanowire
dimers formed by $D=6.2$ nm and $D=9.8$ nm nanowires (see the
legend). Solid and dashed lines show results of classical Drude
calculations for $D=6.2$ nm  (blue) and $D=9.8$ nm (red) dimers.
Dashed lines: separation $S$ is measured between the jellium
edges. Solid lines: separation $S$ is measured between the
centroids of the induced charges. For more details see the main
text.
(c) Energy of the dipole plasmon resonance as function of the
junction width $S$. Dots: TDDFT results obtained for $D=9.8$ nm
nanowire dimer. Solid and dashed red lines show results of
classical Drude calculations [same as (b)]. Dashed and dotted gray
lines show results of NLHD calculations. Dashed line: separation
$S$ is measured between the jellium edges. Dotted line: separation
$S$ is measured between the centroids of the induced charges.
(d) Same as (b), but results are shown as function of the scaled
separation $S/R_{\rm{cl}}$, where $R_{\rm{cl}}=D/2$ is the
nanowire radius.
}
\label{Fig:Ruler}
\end{figure}

With results shown in the previous Subsection, tunneling
determines the optical properties of the system at small
junction widths. For large width $S$ tunneling is absent and
nonlocal dynamic screening influences the optical properties.
Similar to the case of the individual nanowire, the major
nonlocal effect is the red shift of the frequency of the
bonding dipole plasmon as compared to the classical
prediction. However, for the dimer, this effect is much
stronger, and does not disappear with increasing radius
of nanoparticle.

Fig.~\ref{Fig:Ruler} presents the analysis of the role played
by dynamic screening in determining the frequency of the bonding
dipole plasmon. Panel (a) gives schematic representation of the
location of plasmon induced screening charges in the junction
of width $S$. Within the local classical approach the screening
charges are localised at geometrical surfaces of the cylinders
separated by $S$. We recall that we use the definition
of the geometrical surface such that it coincides with jellium
edge. Within TDDFT, the real part of the centroid of the induced 
charge density $Re[d(\omega)]$ corresponds to the position of 
the plasmon-induced surface charges with respect to the jellium 
edge of each cylinder \cite{Apell1982,Liebsch1993,Feibelman1982,
Liebsch1987,Appel1984,Liebsch1995,Tsuei1990}. For the present 
case of sodium nanowires, $Re[d(\omega)]$ is positive 
$\approx 0.9$~{\AA} in the frequency range of 
interest \cite{Liebsch1987}. This means that the screening 
charge is shifted by $0.9$~{\AA} into the vacuum and located 
at $\approx 3$~{\AA} outside the surface atomic plane of the 
nanoparticle. The distance between the plasmon induced charges 
across the junction is thus $\Sigma=S-2 Re[d(\omega)]$. As to 
the NLHD approach, with present choice of the non--locality 
parameter $\beta$ it places the screening charges at 
$\delta=1$~{\AA} inside the geometrical surface. Thus, within 
the NLHD approach the screening charges are separated by the 
distance $S+2 \delta$.

This insight provides an explanation for the junction width 
dependence of the DP modes shown in Fig.~\ref{Fig:Ruler}(b)
for two different nanowire dimers. While the overall agreement
between the classical Drude (dashed lines) and the TDDFT (dots)
calculations is good, the TDDFT result is systematically
redshifted. The difference is increasing with decreasing $S$
indicating that the classical calculations underestimate the
coupling between the dipole plasmons localized on each nanowire.
Our finding that the plasmon-induced charges in the TDDFT
calculations are outside the nanoparticle surfaces explains 
this effect. Indeed, the actual energy of the
DP is given by the interaction between the screening charges
across the junction. Thus, the TDDFT data obtained for
junction width $S$ should be compared with classical results
calculated for $\Sigma=S-2 Re[d(\omega)]$.
If the DP dispersion is plotted as a function of effective
junction width given by the distance between plasmon induced
screening charges, the agreement between TDDFT and the
classical simulations becomes excellent.
We note in passing that the step-structure of the TDDFT data
for the smaller $D=6.2$~nm dimer steams from the strong
coupling between plasmon and electron-hole pair excitations.

A similar analysis can be performed for the NLHD model
calculations shown in Fig.~\ref{Fig:Ruler}(c)
for the case of $D=9.8$~nm nanowire dimer. The NLHD
results appear at even higher frequencies than the classical
Drude calculations, in even stronger disagreement with the
TDDFT than the classical model. The explanation for this is 
that NLHD artificially places the plasmon--induced screening 
charges at a position  $\sim \delta=1$~{\AA} inside the metal
surface. Thus, the separation between screening charges 
$S+2\delta$ in the NLHD model is larger than separation $S$ 
in pure local theory, and too large compared to the quantum 
result $\Sigma=S-2 Re[d(\omega)]$. This decreases the 
interaction and results in blue shifted DP modes compared 
to the classical model and TDDFT. If the DP dispersion is 
plotted as a function of effective junction width, agreement 
between NLHD, classical Drude, and TDDFT results is obtained.

The above results show that the TDDFT dispersion of the
DP can be fully retrieved with local classical calculations.
The main issue here is the size of the junction for which
the calculations have to be performed. For geometry of 
the nanostructure, it seems convenient to define the surfaces
of the objects as given by the jellium edges. In this 
case the width $S=0$ of the junction would correspond to 
continuous solid formed at the contact point, which 
is physically sound. On the other hand, classical 
calculations have to be performed for the effective junction 
width given by the actual separation between the screening 
charges. These conclusions have direct implications for the 
ultimate limit of resolution of plasmonic 
rulers.\cite{Hill2012}

For the free-electron Na surface, the screening charges are
located at approximately $0.9$~{\AA} ($3$~{\AA}) outside
the jellium edge (surface atomic layer) meaning that for a
Na--Na junction, the effective junction width would be
$1.8$~{\AA} ($6$~{\AA}) smaller than the physical junction 
width measured between the jellium edges (surface atomic 
planes). This conclusion holds not only for the nanowire case 
considered here, but for any junctions between Na surfaces.
For silver and gold, analysis of the data on the blue shift
of the dipole plasmon resonance in small clusters
\cite{Liebsch1993,Tigges1993,Scholl2013,gold1,gold2} with
Eq.~\ref{Eq:RedShift} places the effective screening charges 
inside the jellium edge at $1.5 \div 0.85$~{\AA} for silver, 
and at approximately $1.5 $~{\AA} for gold. Therefore, 
for an Ag--Ag (Au--Au) junctions, the effective junction width 
would be by $1.7 \div 3$~{\AA} ($\sim 3 $~{\AA}) larger than 
the physical junction width measured between the jellium 
edges and close to the junction width masured between the 
surface atomic planes.

The use of the plasmon ruler relies on the universal
dependence of the DP frequency on the scaled separation
\cite{Gunnarsson2005,Jain2007}. In Fig.~\ref{Fig:Ruler}(d)
we show the TDDFT and classical results for the DP frequency
of the $D=6.2$ nm and $D=9.8$ nm nanowire dimers as function
of the scaled separation $S/R_{\rm{cl}}$.
The TDDFT data for both nanowire dimer sizes nearly falls
on the unique curve provided that the separations $S$
are sufficiently large that no tunneling occurs. This 
holds for $S$ larger than typically 2 lattice constants
between surface atomic planes, which sets lower bound for the
distances that can be actually measured with plasmon ruler.

\subsection{Summary and Conclusions}

In conclusion, we have presented fully quantum mechanical
study of the optical response of the plasmonic dimer formed
by realistic size cylindrical nanowires. This system is also
representative of interacting nanorods and is of relevance
for SERS, plasmon ruler, and plasmon transport applications.
Translational invariance allows to apply the time--dependent
density functional theory for the plasmonic dimer of
largest size considered so far in quantum calculations.

We considered nanowires made of Sodium which is a prototype
for free-electron metal so that jellium model applies.
The free-electron character of Na valence electrons implies
that its permittivity can be well described with Drude
model, as well as it is consistent with approximations behind
the hydrodynamic approach to model the nonlocal character of
the dielectric function. Thus, this is the system of choice
allowing to set the full quantum TDDFT benchmark results, and
to use these results to test different theoretical approaches
addressing plasmonic response of the strongly coupled objects.
This was one of the central goals of the present work.

We have found that for the small junction widths, the optical
response is determined by the  quantum tunneling of conduction
electrons across the potential barrier separating the nanowires.
A decreasing junction size leads to progressive attenuation
of the plasmon modes of separated nanowires and the emergence
of charge transfer plasmon modes of conductively coupled dimer.
As this happens, the fields in the middle of the junction
are screened. The maximal field enhancement moves from
the middle to the external regions of the junction.
In this distance range the classical
local Drude and nonlocal hydrodynamic model descriptions fail
since they do not account for tunnelling. At variance, the
quantum corrected model reproduces the TDDFT findings on a
semi--quantitative level.

For large separations $S$, the tunneling is negligible and 
the overall agreement between  TDDFT, classical and the
QCM results is good. Thus the QCM performs well over the
entire range of separations studied here. For large $S$
the agreement between the classical and TDDFT results
can be further improved by taking into account the shift of
the plasmon-induced charge density with respect
to the geometrical surface of the nanoparticle.
The latter can be defined as the jellium edge as in the
present work, or as the top most atomic layer at the surface.
The effective junction width is then given by the separation
of plasmon indicted charges at the opposite sides of the
junction. Indeed, this is  the interaction between these
charges that determines the hybridization and energies of
the modes. Introducing a simple distance correction
into the classical calculations allows a full account of this
non-local effect. This result has implications for the plasmon
ruler concept and shows that care should be taken with
respect to the definition of the separation which is actually
measured.


We hope that the results presented here contribute to
the understanding of the role of quantum nonlocal
effects in strongly coupled plasmonic systems, and
help in elaborating efficient theoretical approaches
for their prediction.

ACKNOWLEDGEMENTS We thank Guiseppe Toscano for providing
NPL extension to Comsol 4.2a RF Module used in our
nonlocal calculations as well as for his kind assistance.
J.A. acknowledges financial
support from the Department of Industry of the Basque Government
through the ETORTEK project nanoiker, the Spanish Ministerio de
Ciencia e Innovaci\'{o}on through Project No. FIS2010-
19609-C02-01 and Project No. EUI2008-03816 CUBIHOLE
from the Internationalization Program. P.N. Acknowledges support
from the Robert A.
Welch Foundation (C-1222) and the Office of Naval Research
(N00014-10-1-0989).\\



\end{document}